\documentclass[twocolumn,epjc3]{svjour3}          

\RequirePackage[T1]{fontenc}

\RequirePackage[pdftex]{graphicx}
\RequirePackage{mathptmx}      
\RequirePackage{flushend}
\RequirePackage[numbers,sort&compress]{natbib}
\RequirePackage[colorlinks,citecolor=blue,urlcolor=blue,linkcolor=blue]{hyperref}
\RequirePackage{microtype}
\journalname{Eur. Phys. J. C}

\begin{document}

\title{Constraint on the chemical potentials of hydrogen and proton in recombination}


\author{L. L. Sales\thanksref{e1,addr1}
        \and
        F. C. Carvalho\thanksref{e2,addr1} 
        \and 
        H. T. C. M. Souza\thanksref{e4,addr3} 
}

\thankstext{e1}{lazarosales@alu.uern.br}
\thankstext{e2}{fabiocabral@uern.br}
\thankstext{e4}{hidalyn.souza@ufersa.edu.br}

\institute{Departamento de F\'\i sica, Universidade do Estado do Rio Grande do Norte, 59610--210, Mossor\'o -- RN, Brazil\label{addr1}
          \and
          Departamento de Ci\^encias Exatas e Naturais, Universidade Federal Rural do Semi--\'Arido, 59900--000, Pau dos Ferros -- RN, Brazil\label{addr3}
}

\date{Received: date / Accepted: date}

\maketitle

\begin{abstract}
	
	In this paper, we revisit the hydrogen recombination history from a novel perspective: the evolution of chemical potentials. We derive expressions for the chemical potentials, which depend on the thermal bath temperature and the ionization degree of the universe. Our main finding reveals a constraint between the chemical potentials of hydrogen and proton at $z\approx 1200$ when the free electron fraction is $X_e\approx 1/3$. Furthermore, we present important data on the chemical potentials during recombination, highlighting the differences between the predictions of the Peebles' and CosmoRec code solutions. Finally, we discuss a particular case related to the chemical potential of hydrogen.
	\keywords{Early universe \and Atomic hydrogen recombination \and Chemical potential}
\end{abstract}

\section{Introduction} 

The chemical potential is an abstract measure that relates to the propensity for physical or chemical transformation of a species \cite{gibbs1878equilibrium}. In the field of thermodynamics, the chemical potential of a species stands for the energy that can either be absorbed or released due to a change in the particle number resulting from a chemical reaction or phase transition. This physical quantity is defined as the rate of change of the free energy of a thermodynamic system concerning the alteration in the number of atoms or molecules of the species introduced to the system \cite{atkins2014atkins}.
 
Recombination is a crucial stage in the thermal history of the universe, during which electrons and protons combine to form the neutral hydrogen atom. As the universe expanded and cooled, the primordial plasma underwent a process of neutralization, giving rise to the first hydrogen atoms. At the end of this era, known as the last scattering surface, nearly 99\% of the matter was combined with a significant amount of neutral hydrogen \cite{peebles1993principles}. With the matter in this state almost entirely neutral, the remaining photons traveled freely. They dominated the energy content beyond our Milky Way galaxy, manifesting as the Cosmic Microwave Background Radiation (CMBR) that was first detected by radio astronomers Penzias and Wilson in 1965 \cite{penzias1965measurement}.

One systematic way of understanding the evolution of physical processes throughout the thermal history of the universe is to study it by the Boltzmann equation for annihilation. The reduced form of this equation is given by \cite{dodelson2003modern}
\begin{equation} 
	\frac{df}{dt} = C[f] \;,
\end{equation} 
with $f$ representing the distribution function and $C[f]$ referring to a functional containing all possible collision terms between species (fermions or bosons). In the case of cosmological recombination, with some simplifications in $C[f]$, the equilibrium condition, known as the Saha equation, is established.  

Saha's ionization equation is built under chemical equilibrium and assumes that recombination occurs directly from the ground state. However, since the recombination history is characterized by physical processes that are out of equilibrium, the Saha equation is an inadequate tool for fully describing these processes. In contrast, Peebles' approach to recombination considers the exact solution of the Boltzmann equation, which takes into account non-equilibrium states. Peebles assumes that recombination occurs through the capture of electrons in excited states, which then cascade down to the ground state via the Lyman-$\alpha$ or {2s}-{1s} two-photon transition. Peebles's final expression includes the probability of an excited hydrogen atom decaying via two-photon emission \cite{peebles1968recombination}, thereby releasing photons with insufficient energy to ionize the newly formed hydrogen atoms.
 
This paper focuses on describing the evolution of primordial chemical potentials, particularly during hydrogen recombination in the early universe. Our main objective is to analyze recombination from a new perspective, in order to identify results that are not predicted by traditional approaches. To do so, we shall derive the chemical potentials as a function of the thermal bath temperature and the ionization degree.

The paper is organized as follows. Details of Saha and Peebles' approaches are presented in Section \ref{recombination}. In Section \ref{chemical_potencials}, we show how primordial chemical potentials are ascertained. A special case is discussed in Section \ref{special_case}. Lastly, the conclusion of the paper is presented in Section \ref{conclusion}. 

\section{Description of recombination} \label{recombination}

\subsection{Saha's approach}

Developed by the Indian astrophysicist Meghnad Saha in 1920 \cite{saha1920liii}, the Saha equation plays an important role in understanding several physical processes in the early universe. The expression relates the ionization state of a gas in thermal equilibrium to the temperature and pressure. In order to start Saha's approach to recombination, let us consider the following physical reaction:  
\begin{equation} \label{pf}
	e^{-}+p\leftrightarrow H+\gamma \;,
\end{equation}
where $e^{-}$, $p$, $H$, and $\gamma$ stand for the electron, proton, neutral hydrogen, and photon, respectively. 

Assuming that the reaction (\ref{pf}) remains in equilibrium, the thermal ionization equation can be conveniently written as\footnote{This equilibrium condition considers $n_{\gamma}=n_{\gamma}^{(0)}$ since the photon chemical potential is assumed to be zero (i.e., $\mu_{e}+\mu_{p}=\mu_{H}$). This is the standard assumption and is the case if photons follow a Planck spectrum. It also needs that Compton and double Compton processes are in equilibrium.}
\begin{equation} 
	\frac{n_{e}n_{p}}{n_{H}} = \frac{n_{e}^{(0)}n_{p}^{(0)}}{n_{H}^{(0)}}\;,
\end{equation}
and each species $i$ comes from the integral over momentum space 
\begin{equation} 
	n_{i} = g_{i}e^{\beta\mu_{i}}\int \frac{d^{3}p}{(2\pi\hbar)^{3}}e^{-\beta E_{i}}\;,
\end{equation}
and
\begin{equation}
	n_{i}^{(0)} \equiv g_{i}\int \frac{d^{3}p}{(2\pi\hbar)^{3}}e^{-\beta E_{i}}\;.
\end{equation}
The superscript ``($0$)'' means that the particle number density is defined at equilibrium, i.e., by configuring the chemical potential equal to zero \cite{dodelson2003modern}. In equations above $\beta=1/k_{B}T$, where $k_{B}$ and $T$ are the Boltzmann constant and the thermal bath temperature, respectively. Also, $\mu_{i}$ is the chemical potential of species $i$ and $\hbar$ the reduced Planck's constant. The factor $g_{i}$ is the spin states number and $E_{i}$ corresponds to the energy for nonrelativistic particles ($m_{i}c^{2}\gg k_{B}T$).   

The Saha equation for recombination is given by \cite{dodelson2003modern,peebles1993principles,weinberg2008cosmology}
\begin{equation} \label{saha_eq}
	\frac{X_{e}^{2}}{1-X_{e}} = \frac{1}{n_{b}}\left(\frac{m_{e}k_{B}T}{2\pi\hbar^{2}}\right)^{3/2}e^{-\epsilon_{0}/k_{B}T}~,
\end{equation}
where $\epsilon_{0}=13.6~\mathrm{eV}$ represents the ground state binding energy of the hydrogen atom, and the free electron fraction is defined as follows:
\begin{equation} \label{ioniz_fraction}
	X_{e} = \frac{n_{e}}{n_{b}}\;,
\end{equation} 
with $n_{b}=n_{p}+n_{H}$ being the baryons density\footnote{In hydrogen recombination, charge neutrality demands that $n_p = n_e$ as long as helium is neglected.}, ascertained from baryon to photon ratio by 
\begin{equation}
	n_{b}=1.12\times 10^{-5}\Omega_{b0}h^{2}(1+z)^{3}~\mathrm{cm}^{-3} \;.
\end{equation}
In the equation above, $\Omega_{b0}$ is the baryonic matter density parameter at the present time and $h$ the reduced Hubble constant. 

Saha's equation accounts for recombination directly from the ground state. However, during this process, highly energetic photons are produced, which implies that newly formed neutral hydrogen atoms become ionized. As a result, as $X_{e}$ decreases, the recombination rate decreases, making it more difficult to maintain equilibrium \cite{dodelson2003modern, peebles1993principles}. This leads to a rapid decrease in the ionization degree, reaching zero quickly. Nevertheless, an asymptotic behavior around $10^{-4}$ is expected and necessary for the emergence of the next generation of bound systems, namely molecules \cite{peebles2009finding}. Hence, the Saha equation is only useful for estimating outcomes while maintaining chemical equilibrium. Still, as we will conclude later, this condition will prove to be an exciting indicator for constraining the beginning of the atomic recombination of hydrogen to $z<1400$. 

\subsection{Peebles' approach}

The ionization rate informs us about how the recombination epoch progresses. We use the recombination rate found by Peebles \cite{peebles1993principles,peebles1968recombination}, which takes into account the expansion of the universe, the photo-ionization due to the CMBR, and the photons that arise from the recombination process:
\begin{equation} \label{irate}
	\frac{dX_{e}}{dt} = \left[\beta_{e}(1-X_{e})e^{-[(B_{1}-B_{2})/k_{B}T]} - a_\mathrm{rec}X_{e}^{2}n_{b}\right]C \;.
\end{equation}
In this expression, $B_{1}$ and $B_{2}$ are the binding energies of the ground state and the first excited state of the hydrogen atom, respectively. The parameter $a_\mathrm{rec}$ is the recombination coefficient given by
\begin{equation} 
	a_\mathrm{rec} \equiv \langle\sigma v\rangle = 2.84 \times 10^{-11}T_{m}^{-1/2}~\mathrm{cm}^{3}~\mathrm{s}^{-1} \;,
\end{equation} 
where $T_m$ is the matter temperature, ruled by the equation
\begin{eqnarray} \nonumber
	\frac{dT_m}{dt} = T_m\left[ -2\frac{\dot{a}}{a} - \frac{\dot{X}_e}{3(1+X_e)}\right] -\frac{8}{3}\frac{\sigma_{T}bT^{4}X_e}{m_{e}c(1+X_e)}(T_m-T)~. \\  
\end{eqnarray}

The photo-ionization rate reads as
\begin{equation} 
	\beta_{e} = a_\mathrm{rec}\left(\frac{m_{e}k_{B}T}{2\pi\hbar^{2}}\right)^{3/2}e^{-(B_{2}/k_{B}T)} \;.
\end{equation} 
Besides, the factor $C$ is the probability of an excited hydrogen atom decay via two photons emission ${2s}$-${1s}$, given by
\begin{equation} 
	C = \frac{\Lambda_{2s, 1s}}{\Lambda_{2s, 1s}+\beta_{e}} \;,
\end{equation}
where $\Lambda_{2s, 1s} = 8.23~\mathrm{s}^{-1}$ is the decay rate of two photons from the metastable level $2s$ to the ground state $1s$ \cite{peebles1993principles,jones1985ionisation}. For estimation purposes, we adopt the latest data from the Planck Space Telescope: $\Omega_b=0.0492$, $\Omega_d= 0.264$, $h= 0.674$ and $H_0= 67.4$ km s$^{-1}$ Mpc$^{-1}$ \cite{aghanim2020planck}. In our calculations, we are assuming the flat $\Lambda$CDM model.

\begin{figure}[!h]
	\centering
	\includegraphics[scale=0.09]{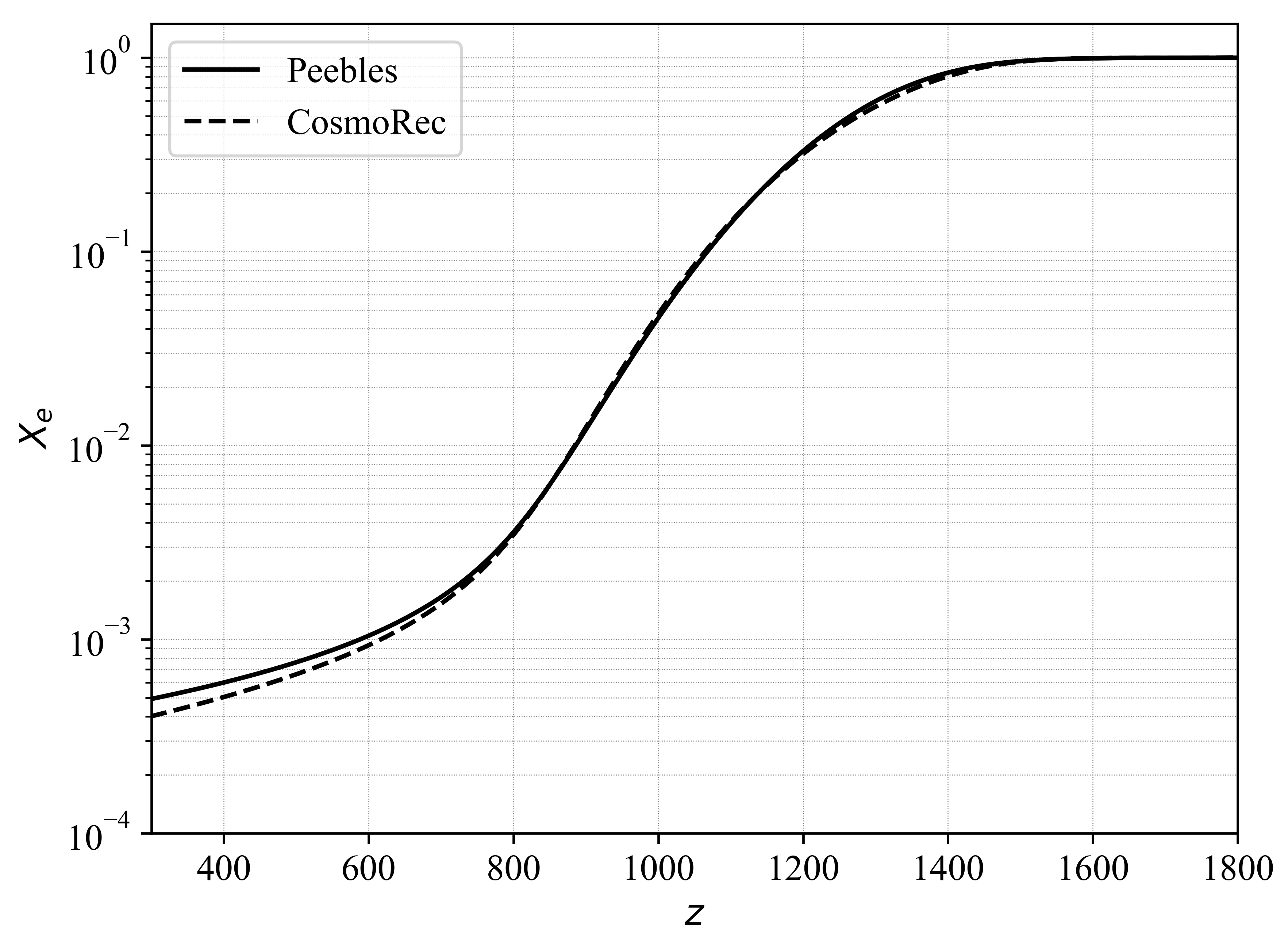}
	\caption{The curves depict the evolution of the ionization degree considering Peebles' model ($X_e=n_e/n_b$) and the result provided by the CosmoRec code ($X_e=n_e/n_H$)} 
	\label{fig.1}
\end{figure}

The most accurate results for the ionization history are given by numerical codes, such as CosmoRec\footnote{Available at \url{http://www.jb.man.ac.uk/~jchluba/Science/CosmoRec/CosmoRec.html}}, which take into account several important corrections to the cosmological recombination history \cite{chluba2011towards}. In this study, we assume that hydrogen recombination begins when $90\%$ ($X_{e}=0.9$) of the universe's matter is still in an ionized state, and ends when $99\%$ has already combined, i.e., when $X_{e}=0.01$. Fig. \ref{fig.1} shows the ionization history during recombination obtained through numerical integration of Eq. (\ref{irate}) and CosmoRec code.  

\noindent
Notice that Peebles' solution agrees reasonably well with the CosmoRec code for $z>800$. In the present study, our focus is solely on the hydrogen recombination history. We assume that helium recombination occurred well before this stage, ensuring that the ratios $n_e/n_b$ and $n_e/n_H$ are equivalent for $X_e<1$. For comparison purposes, Table \ref{table1} displays important numerical results for recombination obtained through both Peebles' approach and the CosmoRec code.

\begin{table}[!h] 
	\begin{center}
		\caption{Some data on the hydrogen recombination epoch. The \emph{beg} and \emph{end} indices denote the beginning and end of recombination, respectively. Additionally, $\Delta z_{\mathrm{rec}}$ represents the duration of recombination in redshift, while the redshift of decoupling ($X_e=0.5$) is denoted by $z_{\mathrm{dec}}$}
		\label{table1} 
		\begin{tabular}{ccccc}
			\hline\noalign{\smallskip}
			Approach & $z_\mathrm{beg}$ & $z_\mathrm{end}$ & $z_\mathrm{dec}$ & $\Delta z_\mathrm{rec}$ \\
			\noalign{\smallskip}\hline\noalign{\smallskip}
			Peebles & 1444.9 & 887.1 & 1264.9 & 557.8 \\
			CosmoRec & 1453.4 & 885.3 & 1277.8 & 568.1 \\
			\noalign{\smallskip}\hline
		\end{tabular}
	\end{center} 
\end{table}

\section{Evolution of primordial chemical potentials} \label{chemical_potencials}

In this Section, we shall ascertain the chemical potentials of species $i$ as a function of the ionization degree and the thermal bath temperature. This will be performed through the utilization of the definition of fugacity. We shall assess the evolution of primordial chemical potentials by employing both Peebles' and CosmoRec solutions. In order to express all energies and chemical potentials in terms of electron-volt (eV), we adopt the natural units system with $c=1$. To commence this evaluation, first consider the following definition:
\begin{equation} \label{density_ratio}
	\frac{n_{i}}{n_{i}^{(0)}} = e^{\beta \mu_{i}} \;. 
\end{equation}
Using the ionization fraction given by Eq. (\ref{ioniz_fraction}) together with Eq. (\ref{density_ratio}), we find the electron chemical potential
\begin{equation} \label{mie}
	\mu_{e} = k_{B}T\left[\ln(n_{b}X_{e}) - \ln(n_{e}^{(0)})\right] \;, 
\end{equation}  
where 
\begin{equation}
	n_{e}^{(0)} = 2\left(\frac{m_{e}k_{B}T}{2\pi\hbar^{2}}\right)^{3/2} e^{-\beta m_{e}} \;.
\end{equation}

Employing the universe neutrality property ($n_{e}=n_{p}$) in Eq. (\ref{ioniz_fraction}), the proton chemical potential is obtained from  
\begin{equation} \label{mip}
	\mu_{p} = k_{B}T\left[\ln(n_{b}X_{e}) - \ln(n_{p}^{(0)})\right] \;, 
\end{equation}  
where 
\begin{equation}
	n_{p}^{(0)} = 2\left(\frac{m_{p}k_{B}T}{2\pi\hbar^{2}}\right)^{3/2} e^{-\beta m_{p}} \;.
\end{equation}

In order to determine the hydrogen chemical potential, we resort the definition of baryons density ($n_{b}=n_{e} + n_{H}$) in Eq. (\ref{density_ratio}), so that
\begin{equation} \label{eH_ratio}
	\frac{n_{e}}{n_{H}^{(0)}} = \frac{n_{b}}{n_{H}^{(0)}} - e^{\beta \mu_{H}} \;, 
\end{equation}
where 
\begin{equation}
	n_{H}^{(0)} = 4\left(\frac{m_{H}k_{B}T}{2\pi\hbar^{2}}\right)^{3/2} e^{-\beta m_{H}} \;.
\end{equation}
Using Eq. (\ref{eH_ratio}) in Eq. (\ref{ioniz_fraction}), we get the hydrogen chemical potential
\begin{equation} \label{mih}
	\mu_{H} = k_{B}T\left[\ln(1-X_{e}) + \ln \alpha\right] \;\;,
\end{equation}
with $\alpha=n_{b}/n_{H}^{(0)}$.

For the sake of simplicity, equations (\ref{mie}), (\ref{mip}) and (\ref{mih}) can be written respectively as
	\begin{eqnarray} \label{mie1} 
		\beta(\mu_{e}-m_{e}) = \ln(X_{e}n_{b}) + \ln\left( \frac{h^{3}}{2}\right)  - \frac{3}{2}\ln(2\pi m_{e}k_{B}T)\;,
	\end{eqnarray}
	\begin{eqnarray} \label{mip1} 
		\beta(\mu_{p}-m_{p}) = \ln(X_{e}n_{b}) + \ln\left( \frac{h^{3}}{2}\right)  - \frac{3}{2}\ln(2\pi m_{p}k_{B}T) \;,
	\end{eqnarray}
	and
	\begin{eqnarray} \label{mih1} \nonumber
		\beta(\mu_{H}-m_{H}) &=& \ln((1-X_{e})n_{b}) + \ln\left( \frac{h^{3}}{4}\right) \\
		&-& \frac{3}{2}\ln(2\pi m_{H}k_{B}T)\;.
	\end{eqnarray}
Based on the equations presented above, we can investigate how the chemical potentials of the species evolve throughout the hydrogen recombination history. It is important to note that these potentials depend on the thermal bath temperature and the free electron fraction. 

\begin{figure}[!h]
	\centering
	\includegraphics[scale=0.09]{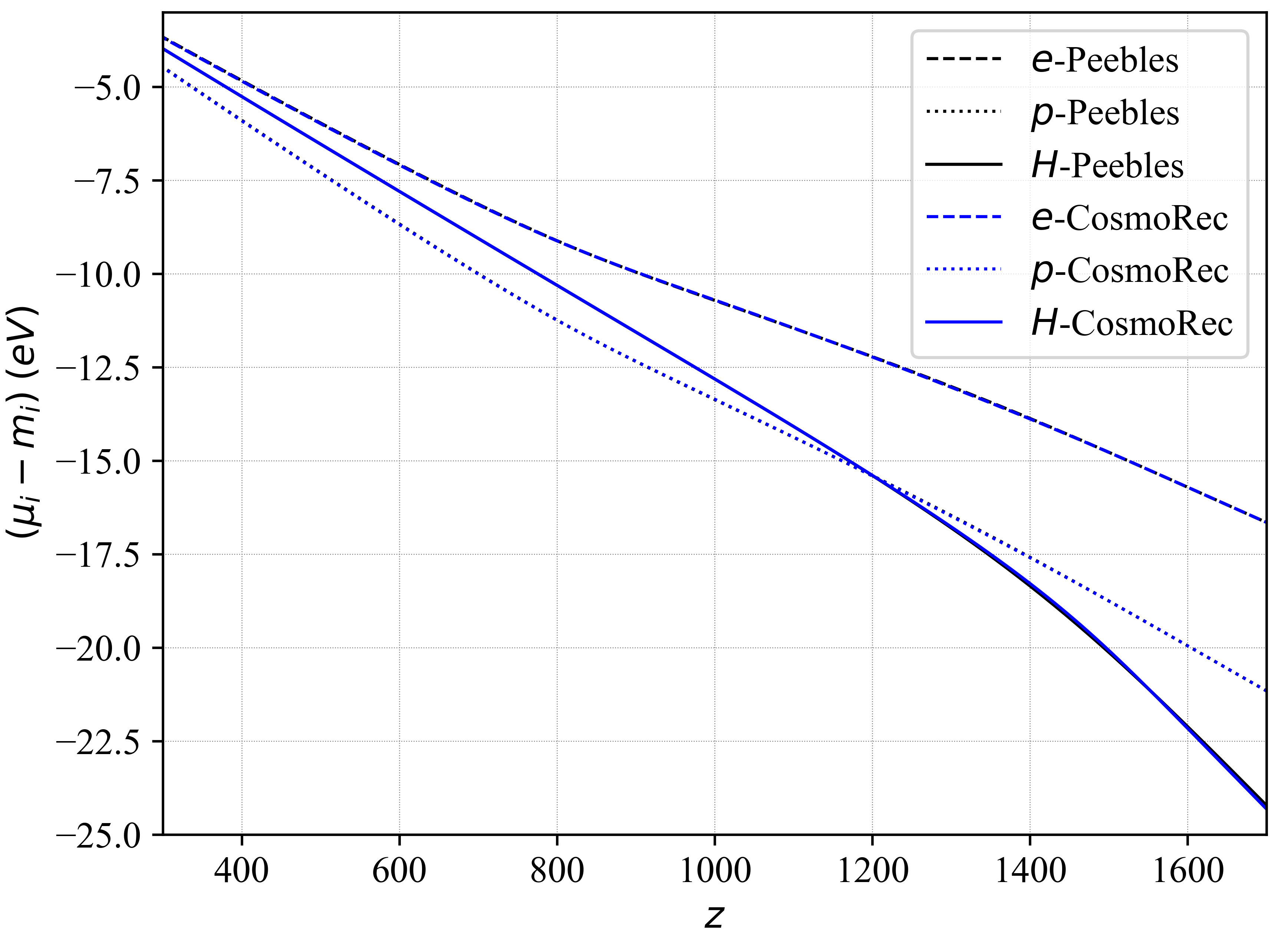}
	\caption{Evolution of chemical potentials during recombination. The dashed, dotted, and solid lines stand for the electron, proton, and hydrogen, respectively. The color black represents Peebles' solution, and the color blue denotes the CosmoRec solution}
	\label{fig.2}
\end{figure}

To incorporate the results of the Peebles' and CosmoRec approaches into our analysis, we use the values of $X_e$ provided by the solution of Eq. (\ref{irate}) and the CosmoRec code in equations (\ref{mie1}), (\ref{mip1}) and (\ref{mih1}). Fig. \ref{fig.2} depicts the evolution of chemical potentials as a function of redshift during the hydrogen recombination history. Note that the chemical potential lines generated via Peebles' solution practically overlap with those yielded through the CosmoRec code. As shown in Fig. \ref{fig.1}, for $z<800$, there is a slight discrepancy between the forecasts of the two approaches. However, from the perspective of chemical potentials, there is no significant difference observed, even for redshifts as low as $z<800$. It is worth stressing that no proton or electron is created and their particle chemical potentials are conserved at these late stages.  

The main finding of our approach is underscored in Fig. \ref{fig.2}, which shows that during the hydrogen recombination history, the curves for $\mu_{H}-m_{H}$ and $\mu_{p}-m_{p}$ intersect for both Peebles' and CosmoRec solutions. This result imposes a noteworthy constraint on the chemical potentials of hydrogen and proton, which can be expressed by
\begin{equation} 
	\mu_{H} - \mu_{p} = 0.511\;\mathrm{MeV}\;,
\end{equation}
where we use the constants $m_H=938.783\;\mathrm{MeV}$ and $m_p=938.272\;\mathrm{MeV}$. By matching equations (\ref{mip1}) and (\ref{mih1}), we might be able to show that the intersection occurs when the ionization degree of the universe is 
\begin{equation}
	X_e = \frac{1}{1+2\left(\frac{m_H}{m_p}\right)^{3/2}}~.
\end{equation}
Considering $m_{p} \approx m_{H}$, we obtain $X_{e}\approx 1/3$, corresponding to a redshift of about $1200$. In terms of cosmic time, this corresponds to when the universe was approximately 413,890 years old. It should be mentioned that the traditional approach to the hydrogen recombination does not predict this result. 

Table \ref{table2} presents important data for primordial chemical potentials during the hydrogen recombination history using Peebles' and CosmoRec code solutions. The first column displays the two methods utilized. Columns two to seven show the electron, proton, and hydrogen values of $y_{i}^{j}$, where we define $y_{i}^{j} = \mu_{i}-m_{i }$ with $j=\mathrm{beg},\:\mathrm{end},\; \mathrm{dec}$. The eighth to last column shows the results for the decoupling. The results presented in Table \ref{table2} only confirm what we had already observed in Fig. \ref{fig.2}: the difference between the predictions according to Peebles and CosmoRec for chemical potentials is very small. 

\begin{table}[!h] 
	\begin{center}
		\caption{Some important outcomes on chemical potentials in hydrogen recombination}
		\label{table2}
		\resizebox{8.5cm}{!}{%
			\begin{tabular}{cccccccccc}
				\hline\noalign{\smallskip}
				Approach & $y_{e}^\mathrm{beg}$ & $y_{e}^\mathrm{end}$ & $y_{p}^\mathrm{beg}$ & $y_{p}^\mathrm{end}$ & $y_{H}^\mathrm{beg}$ & $y_{H}^\mathrm{end}$ & $y_{e}^\mathrm{dec}$ & $y_{p}^\mathrm{dec}$ & $y_{H}^\mathrm{dec}$ \\ 
				\noalign{\smallskip}\hline\noalign{\smallskip}
				Peebles & -14.26 & -9.85 & -18.09 & -12.20 & -19.07 & -11.39 & -12.72 & -16.07 & -16.28 \\
				CosmoRec & -14.35 & -9.83 & -18.20 & -12.18 & -19.18 & -11.37 & -12.85 & -16.23 & -16.44 \\ 
				\noalign{\smallskip}\hline 
		\end{tabular}}
	\end{center}
\end{table}

\section{A special case: $\mu_{H}-m_{H}$} \label{special_case}

Examining the hydrogen recombination history through the lens of chemical potentials offers distinct advantages, as it enables the exploration of aspects that may not be perceptible using traditional approaches. In this section, we discuss the behavior of the chemical potential of hydrogen regarding equilibrium and non-equilibrium states. For equilibrium states, we consider Saha's solution for $X_e$, and for non-equilibrium states, we employ $X_e$ from the CosmoRec code solution. 

\begin{figure}[!h]
	\centering
	\includegraphics[scale=0.09]{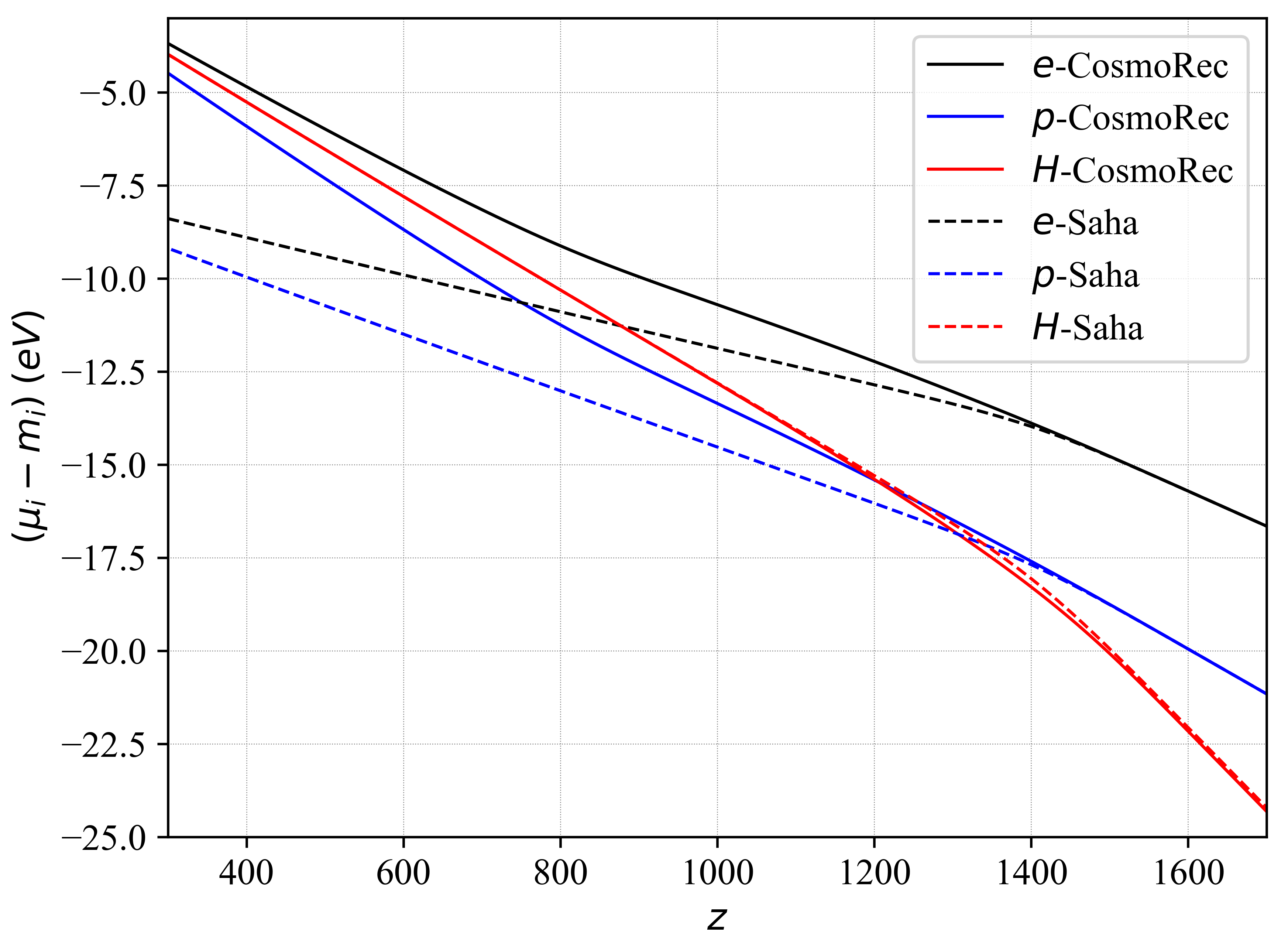}
	\caption{Behavior of chemical potentials for equilibrium and non-equilibrium states. The CosmoRec solution is represented by solid lines, while Saha's solution is depicted by dashed lines. Electron, proton, and hydrogen are represented by black, blue, and red colors, respectively}
	\label{fig.3}
\end{figure}

Fig. \ref{fig.3} portrays the evolution of chemical potentials under both scenarios. Notably, the curves for the hydrogen chemical potential exhibit reasonable agreement, particularly for $z<1200$. As is widely recognized, Saha's solution diverges from the CosmoRec solution at the onset of recombination, so that $X_e$ quickly goes to zero. Thus, one might expect the graph of $\mu_{H}-m_{H}$ yielded by Saha's solution to diverge from the one provided by the CosmoRec solution soon after recombination begins, mainly towards its end. However, as depicted in Fig. \ref{fig.3}, this is not the case. Therefore, we can establish the following result:
\begin{equation}
	(\mu_{H}-m_{H})_\mathrm{CosmoRec} \sim (\mu_{H}-m_{H})_\mathrm{Saha}~.
\end{equation}
 
\noindent
This finding suggests that the quantity $\mu_{H}-m_{H}$ cannot discern between equilibrium and non-equilibrium states. On the other hand, this does not turn out for the electron and proton chemical potentials. Specifically, $\mu_{e}-m_{e}$ and $\mu_{p}-m_{p}$ effectively distinguish between equilibrium and non-equilibrium states, particularly for $z<1400$. The point at which the chemical potential curves of the electron and proton begin to diverge identifies the threshold of the transition from equilibrium to non-equilibrium regimes. This threshold is useful for estimating the start of primordial hydrogen recombination.

\section{Conclusion} \label{conclusion} 

In this paper, we have examined the hydrogen recombination history from the perspective of chemical potentials. Our analysis has uncovered a constraint between the hydrogen and proton chemical potentials at $z\approx 1200$ when the free electron fraction is approximately $1/3$. Additionally, we have presented outcomes on the chemical potentials during recombination, highlighting the differences between the Peebles' and CosmoRec code predictions. It was also shown that the measure $\mu_{H}-m_{H}$ does not discern between equilibrium and non-equilibrium states. This finding offers a potential avenue to simplify the description of recombination without resorting directly to kinetic equations. This is relevant and might be further addressed in future studies. In conclusion, our investigation has yielded meaningful results not previously observed via traditional approaches.  

\begin{acknowledgements}
	The authors are grateful to the Brazilian agency CAPES for financial support. FCC was supported by CNPq/FAPERN/PRONEM.
\end{acknowledgements}


\bibliographystyle{spphys}       
\bibliography{mybibfile}

\end{document}